\documentstyle[preprint,aps] {revtex}
\def\et{{\it et al.\ }}
\def\prb{Phys. Rev. B\ }
\def\ebp{empirical bowing parameter}
\def\tb{tight--binding}
\def\obp{optical bowing parameter}
\def\tbp{tight--binding parameter}
\def\be{\begin{equation}}
\def\ee{\end{equation}}
\begin{document}
\draft
\preprint{CIEA-CM:090495}
\title{Electronic structure and band gap composition--dependence 
of the II--VI quaternary alloys}

\author{A. E. Garc\'\i a, A. Zepeda--Navratil, A. Camacho$^{\rm a}$, 
D. Olgu\'\i n, and R. Baquero}
\address{
Departamento de F\'\i sica, Centro de Investigaci\'on y de Estudios 
Avanzados del Instituto Polit\'ecnico Nacional,
AP. 14-740, M\'exico, D.F.,
Mexico 07000 
}
\address{
$^{\rm a}$ Departamento de F\'\i sica,
Universidad de los Andes, A. P. 4976, Santa F\'e de 
Bogot\'a, Colombia.
}
\maketitle
\begin{abstract}
Based on a successful description of II-VI ternary alloys,
which introduces an \ebp\ to the widely used virtual crystal 
approximation, we set up a tight--binding Hamiltonian to describe
the Zn$_{1-y}$Cd$_y$Se$_{1-x}$Te$_x$ 
and Zn$_{.9}$Cd$_{.1}$S$_{.07}$Se$_{.93}$ quaternary alloys.
We just use a formula that 
can be thought as a straightforward generalization of the 
virtual crystal approximation for this case. Our Hamiltonians 
reproduce very well the change in the band gap value with the 
composition observed in recent experimental reports. 
\end{abstract}

\narrowtext

Interest in optical devices that can operate in the visible spectrum range
has motivated a vigorous research in II--VI wide band gap semiconductor 
compounds. 
The main goal in these studies 
is to seek for the independent control of the fundamental parameters of the 
compound: the band gap, the lattice constant and the valence band 
offset. 
The interest has been recently focused on the study of
the ternary and 
quaternary alloys. 
In the quaternary alloys it is the simultaneous substitution of 
the anion and the cation that allows the manufacture of a compound 
with band gap and the 
lattice constant values within certain limits. This procedure is very 
important in optoelectronic engineering. In fact, it could be
produced a binary/quaternary heterojunction with no
stress at the interface and with a predetermined value for the 
band offset \cite{brasil,brasil2,ichino}. 

Recently, Brasil \et\cite{brasil} have reported 
photoconductivity and photoluminiscence experiments on the 
Zn$_{1-y}$Cd$_y$Se$_{1-x}$Te$_x$ system, and they have fitted 
the measured band gap value to a quadratic function of 
the composition; Ichino \et\cite{ichino} have used the 
Zn$_{.9}$Cd$_{.1}$S$_{.07}$Se$_{.93}$
quaternary alloy as an active layer in a laser 
diode operating in the blue--green to the UV spectral region.

In a previous work, we have described
the change of the band gap value for the ternary 
alloys and compared the results with the existing 
experimental data \cite{ternario}.
To describe the ternary alloys we have
used carefully studied tight--binding Hamiltonians for the  
binary compounds \cite{sup}.
We took into account nearest neighbors interactions with 
an orthogonal basis of five atomic orbitals $(sp^3s^*)$ per atom; 
in this approximation the 
$s^*$--state is included to better describe the lower states 
in the conduction band. 
We have used the two center integrals 
approximation in the Slater--Koster language
and we took into account the spin--orbit 
interaction \cite{slater,romeo}. 
The tight--binding method 
describes rather well the valence band region and gives correctly 
the band gap value \cite{harrison}.
To include composition we have
generalized the virtual crystal approximation (VCA) by 
introducing an \ebp. 
As we have found, the VCA model can describe properly the 
behavior of the band gap if we introduce an \ebp\ 
to calculate the ``$s$'' 
on--site \tbp\ of the substituted ion \cite{ternario}.


In the semiconductor compounds the band gap, $E_g$, is given by the difference 
between the minimum of the conduction band, $E_c$, and the maximum 
of the valence band, $E_v$, 
\be
E_g=E_c-E_v
\ee
In particular, for the compounds with direct transition 
the expressions for $E_c$ and $E_v$ at $\Gamma$ , in the tight--binding method,
are: \cite{romeo}
\be
E_c={E_{sa}+E_{sc}\over 2}+\sqrt{\Big({E_{sa}-E_{sc}\over 2}\Big)^2
				+V_{ss}^2}
\ee
and
\be
E_v={E_{pa}+E_{pc}+\lambda_a+\lambda_c\over 2} -\sqrt{
\Big({E_{pa}-E_{pc}+\lambda_a-\lambda_c\over 2}\Big)^2+V_{xx}^2}
\ee
where $E_{\alpha\nu}$ ($\alpha= s,\ p;\ \nu=a,\ c$) are the on--site
\tbp s with $s$ or $p$ character for the anion ($a$) and the cation ($c$), 
$V_{\alpha\alpha}$ are the nearest neighbor interaction parameters, and 
$\lambda_\nu$ the spin--orbit parameters. 

To describe the alloy we have used the VCA model, as we stated above.
In this approximation the \tbp s are given by the weighted averages 
of the corresponding end--point parameters. For a 
ternary alloy AB$_{1-x}$C$_x$, for example, we have:
\be
{\overline E}_{\alpha\alpha'}(x)=xE_{\alpha\alpha'}^{\rm AC}+
(1-x)E_{\alpha\alpha'}^{\rm AB}
\label{cuatro}
\ee
where $E^j$ is the corresponding \tbp\ of the different binary compounds 
and $\alpha, \alpha'$ denote the 
atomic orbitals ($sp^3s^*$). We will use this expression for all the \tbp\ 
but the ``$s$'' on--site ones. 

As it is well known, the VCA model by itself does not describe correctly the 
non--linear behavior of the band gap value in the alloy. Therefore, 
to take into account the non--linear behavior 
of the band gap in the ternary description, 
we have included an \ebp\ in the ``$s$'' on--site \tbp\ of the 
substituted ion. 
This is because the ``$s$'' on--site \tbp\ is responsible for 
the correct energy position of the conduction band in the 
$\Gamma$--point \cite{ternario}, as can be inferred
from eq. (2). In an explicit way, when we have an
anion substitution we introduce an \ebp\ in the ``$s$'' on--site
anion--VCA expression, and the same for the cation 
case. We use the next VCA expression for the ``$s$'' on--site 
\tbp 
\be 
\overline E'_{s\nu}(x,b_\nu)=\overline E_{s\nu}(x)+b_\nu x(1-x),\qquad \nu=a,\ c
\ee
where $\overline E_{s\nu}(x)$ is given by eq. (\ref{cuatro}),
and $b_\nu$ is the \ebp\ appropriate to 
each substitution. We define the \ebp\ as 
\be 
b_\nu=\pm k{\mid E_{s\nu}^1-E_{s\nu}^2\mid^\lambda\over\mid V_{ss}^1-
				V_{ss}^2\mid},
\ee
where
$E_{s\nu}^{1(2)}$ is the ``$s$'' on--site \tbp\ for the compound 1 (2), 
$V_{ss}^{1(2)}$ the corresponding nearest neighbour $s-s$ interaction parameter
for the compound 1 (2), and  we use the sign $+ (-)$ 
for cation (anion) substitution, 
the proportionality constant is taken as $k=1/8$, and 
we take $\lambda$ as our free parameter in order to obtain the 
best fit to the experimental data. For the most of the alloys we 
have used $\lambda=1.75$ \cite{ternario}, and for the S--based
alloys we have used $\lambda=1.0$. 
We will show that these values
for $\lambda$ give us good agreement with experiment. 
The 
values for the empirical bowing parameter, given in Table I, are the same
that we have used in the ternary description (see Ref. \cite{ternario}).
With this Hamiltonians we can calculate the band gap for any composition 
in a straightforward way. 
However, before
to do that we should to extend our approximation in order to incorporate 
properly the quaternary alloys.

In the 
quaternary alloy A$_{1-y}$B$_y$C$_{1-x}$D$_x$ the VCA expression 
for the \tbp s, that is the appropriate extension of eq. (\ref{cuatro}), is
\be
{\overline E}_{\alpha\alpha'}(x,y)=xyE_{\alpha\alpha'}^{\rm BD}+
x(1-y)E_{\alpha\alpha'}^{\rm AD}+(1-x)yE_{\alpha\alpha'}^{\rm BC}
+(1-x)(1-y)E_{\alpha\alpha'}^{\rm AC},
\label{cinco}
\ee
where the $E^j$'s are defined in eq. (\ref{cuatro}).
In these alloys we have both 
substitutions, anion-- as well as cation--substitution.
For that reason we have found natural to extend (in the spirit
of the VCA) our formula, for the ``$s$'' on--site parameters, 
of the ternary alloys as follows:

\noindent For the anion substitution, 
\be 
\overline E_{sa}(x,y,b_a)=\overline E_{sa}(x,y)+x(1-x)(yb_a^{\rm B}
			+(1-y)b_a^{\rm A}),
\label{siete}
\ee
where $\overline E_{sa}(x,y)$ is given by eq. (\ref{cinco}), $b_a^{\rm B}$ 
and $b_a^{\rm A}$ are the \ebp s for the anion substitution
in the BC$_{1-x}$D$_x$ and AC$_{1-x}$D$_x$ systems, respectively.

\noindent For the cation substitution, 
\be 
\overline E_{sc}(x,y,b_c)=\overline E_{sc}(x,y)+y(1-y)(xb_c^{\rm D}+
			(1-x)b_c^{\rm C}),
\label{ocho}
\ee
where $\overline E_{sc}(x,y)$ is given by eq. (\ref{cinco}), $b_c^{\rm D}$
and $b_c^{\rm C}$ are the \ebp s for the cation substitution
in the A$_{1-y}$B$_y$D and A$_{1-y}$B$_y$C systems,
respectively. 
Note that in the end--values of the compositional variable $y$
in eq. (\ref{siete}) (or $x$ in eq. (\ref{ocho})) 
we obtain the appropriate expression 
for the ternary case, eq. (5). Using these expressions, for the ``$s$''
on--site \tbp s, we can calculate the band gap value and the 
electronic structure for any composition of the quaternary alloys.

Table II shows a comparison of our calculation of $E_g(x,y)$
and the 
photoconductivity measurements of Brasil \et\ 
for Zn$_{1-y}$Cd$_y$Se$_{1-x}$Te$_x$ \cite{brasil}.
As can be judged through the study
of the table our calculation agrees well with the measured values.
In general, we reproduce the experimental values within
$2\%$ accuracy.

Fig. 1 shows a 3--D graph for $E_g(x,y)$. The dots are the 
experimental data, the mesh is our calculation. Note that all the 
experimental data are just over the calculate surface. The 
edges are the four ternary
boundaries. As we have noticed, our calculation reproduces well the
known bowing of $E_g$, both in the anion substitution as in the
cation case. In the anion substitution the bowing of $E_g$ is more
noticeably than in the cation one. 
This fact can be inferred from the figure if we take
the projection of $E_g$ on the $X-Z$ plane. In the same way, we
also appreciate the cuasi--linear behavior of $E_g$ in the cation
substitution case. We can see this facts from the Table I as well.
Notice that, in
absolute value, the bowing parameter for the anion substitution
is greater than the cation one. Although our \ebp\ is not
given by the experiment it can be shown, through the \tb\ equations,
that a great \ebp, in absolute value, gives great optical
bowing parameter; and the \obp\ is the measured value
in the experiment (see Ref. \cite{ternario} for details in this fact).

We have calculated the electronic structure for the 
Zn$_{.9}$Cd$_{.1}$S$_{.07}$Se$_{.93}$ quaternary alloy, as well. 
The particular interest for this quaternary alloy is the 
experimental report of Ichino \et\cite{ichino} These authors 
propose the alloy as an active layer in laser diode operating in
the blue--green to ultraviolet spectral region. Fig. 2 shows our 
calculated electronic structure. This electronic structure is representative 
of a semiconductor compound with direct band gap. From the figure we appreciate
that the lowest conduction band shows lesser dispersion than the 
calculated one, using the empirical pseudopotential method (EPM), 
by Feng \et\cite{feng}. However, the general pattern obtained 
for the valence band is the expected one from tight--binding 
calculations. The calculated band gap for this alloy is 2.705 
eV that is in good agreement with the experimental value of 2.73 eV at 
4.2 K given by Ichino \et\cite{ichino}, and better than 
the value of 2.648 eV calculated by Feng \et\cite{feng}
using the EPM. 

It is noteworthy that:
This work is based on a correct
description for the ternary systems.
We believe that this fact gives 
an important support to our method in general and to the potential 
use of this Hamiltonians in further calculations for surfaces, 
interfaces, quantum wells and superlattices of II-VI compound 
systems \cite{futuro}.

In conclusion, we have calculated the changes in the band gap value of 
the Zn$_{1-y}$Cd$_y$Se$_{1-x}$Te$_x$ 
and Zn$_{.9}$Cd$_{.1}$S$_{.07}$Se$_{.93}$ quaternary alloys 
within a tight--binding description. For the pure 
binary compounds, we have used the \tbp s which describe well the 
known band structures. To describe the alloys, we have used the 
virtual crystal approximation reformulated according to reproduce 
the observed non linear behavior of the band gap with the composition.
We have introduced an \ebp\ in the \tbp s of the ``$s$'' atomic 
orbitals which are known to be responsible for the correct energy 
position of the conduction band in the $\Gamma$--point. Our results 
agree well with experimental results. Our Hamiltonians can be used 
as a basis for other calculations that include this kind of 
pseudobinary compounds.

\newpage

\narrowtext
\begin{table}
\caption{Empirical bowing parameters obtained using the eq. (6) 
and the \tbp s given in the Refs. [5,11]
}
\begin{tabular}{ccc}
 Compound      &       b$_a$   &       b$_c$   \\
\tableline
        ZnSe$_{1-x}$Te$_x$          &    --6.964&    $-$      \\
        CdSe$_{1-x}$Te$_x$          &    --0.195&   $-$       \\
        CdS$_{1-x}$Se$_x$           &    --0.136&             \\
        ZnS$_{1-x}$Se$_x$           &    --2.833&             \\
\hline
        Zn$_{1-x}$Cd$_x$Se          &     $-$   &     0.037   \\
        Zn$_{1-x}$Cd$_x$Te          &     $-$   &     0.020   \\
        Zn$_{1-x}$Cd$_x$Se          &     $-$   &     1.349   \\
\end{tabular}
\label{tabla1}
\end{table}

\narrowtext
\begin{table}
\caption{Measured values for the band gap given in
Ref. [1] for the Zn$_{1-y}$Cd$_y$Se$_{1-x}$Te$_x$ quaternary 
alloys are compared 
with our calculation. Band gap values given in eV.
}
\begin{tabular}{cccc}
 $x$ & $y$ & $E_g({\rm PC})^{\rm a}$ & $E_g({\rm theo})^{\rm b}$ \\
\tableline
        0.015 & 0.10  & 2.610 & 2.642 \\
        0.020 & 0.23  & $-$   & 2.453 \\
        0.050 & 0.22  & 2.443 & 2.427 \\
        0.070 & 0.34  & 2.244 & 2.256 \\
        0.080 & 0.28  & 2.362 & 2.316 \\
        0.038 & 0.32  & 2.024 & 2.002 \\
        0.039 & 0.11  & 2.224 & 2.194 \\
        1.000 & 0.31  & 2.049 & 2.099 \\
\end{tabular}
\label{tabla3}
$^{\rm a}$ Values taken from photoconductivity measurements
of Brasil \et Ref. [1]\\
$^{\rm b}$ This work\\
\end{table}

\newpage

\begin{figure}
\caption{3--D representation for the calculated band gap of the 
Zn$_{1-y}$Cd$_y$Se$_{1-x}$Te$_x$ quaternary alloy, as function of the 
composition $(x,y)$, using the tight--binding method and the 
virtual crystal approximation as is proposed in this work.
The points are the experimental data 
of Brasil \et\ [1].}
\label{fig1}
\end{figure}

\begin{figure}
\caption{
Electronic structure for the Zn$_{.9}$Cd$_{.1}$S$_{.07}$Se$_{.93}$
quaternary alloy calculated using our tight--binding parametrization.}
\end{figure}


\begin{references}
\bibitem{brasil} M. J. S. P. Brasil, M. C. Tamargo, R. E. Nahory, H. 
L. Gilchrist, and R. J. Martin, Appl. Phys. Lett. {\bf 59}, 1206 (1991).
\bibitem{brasil2} M. J. S. P. Brasil, R. E. Nahory, 
F. S. Turco-Sandroff, H. L. 
Gilchrist, and R. J. Martin Appl. Phys. Lett. {\bf 58}, 2509 (1991).
\bibitem{ichino} K. Ichino, Y. --H. Wu, Y. Kawakami, S. Fujita,
and S. Fujita, J. Cryst. Growth {\bf 117}, 527 (1992)
\bibitem{ternario} D. Olgu\'\i n, R. de Coss, and R. Baquero, 
to be published.
\bibitem{sup}D. Olgu\'\i n and R. Baquero, \prb\ {\bf 51}, 1681 (1995);
D. Olgu\'\i n and R. Baquero, \prb\ {\bf 50}, 1498 (1994)
\bibitem{slater} J. C. Slater and G. F. Koster, Phys. Rev. {\bf 94}, 
1498 (1954);
D. J. Chadi, \prb\ {\bf 16}, 790 (1977)
\bibitem{romeo} R. Baquero, R. de Coss, and D. Olgu\'\i n, unpublished. 
\bibitem{harrison} 
W. A. Harrison, \prb\ {\bf 24}, 5835 (1981);
K. C. Hass, H. Ehrenreich and B. Velick\'y, Phys. 
Rev. B {\bf 27}, 1088 (1983).
\bibitem{feng} Y. P. Feng, K. L. Teo, M. F. Li, H. C. Poon, C. K. Ong,
and J. B. Xia, Appl. Phys. Lett. {\bf 74}, 3948 (1993)
\bibitem{futuro} D. Olgu\'\i n and R. Baquero, to be send.
\bibitem{doc} D. Olgu\'\i n, Ph. D. Thesis, CINVESTAV--IPN, Mexico 1996.
\end{references}
\end{document}